# Measurement and Analysis of an Online Content Voting Network: A Case Study of Digg

Yingwu Zhu
Seattle University, Seattle, USA
zhuy@seattleu.edu

## ABSTRACT

Emergence of online content voting networks allows users to share and rate content including social news, photos and videos. The basic idea behind online content voting networks is that aggregate user activities (e.g., submitting and rating content) makes high-quality content thrive through the unprecedented scale, high dynamics and divergent quality of user generated content (UGC). To better understand the nature and impact of online content voting networks, we have analyzed Digg, a popular online social news aggregator and rating website. Based on a large amount of data collected, we provide an in-depth study of Digg. In particular, we study structural properties of Digg social network, impact of social network properties on user digging activities and vice versa, distribution of user diggs, content promotion, and information filtering. We also provide insight into design of content promotion algorithms and recommendation-assisted content discovery. Overall, we believe that the results presented in this paper are crucial in understanding online content rating networks.

## 1. INTRODUCTION

The advent of user generated content (UGC) has dramatically reshaped the landscape of Internet, shifting the role of many websites from creating online content to providing facilities for Internet users to publish their own content and empowering the role of Internet users from content consumers only to content publishers, referees and consumers. Online content rating networks are among such websites. Example systems include YouTube [8], Flickr [3] and Digg [2] where users share and rate videos, photos and news, respectively.

Driven by unprecedented scale, high dynamics and divergent quality of UGC, online content rating networks are creating new viewing patterns, information filtering techniques, content discovery channels and social interactions. One striking feature in online content rating networks is that popularity and availability of content are driven by users' participation, i.e., rating or voting on content. For example, Digg and Flickr have a front page that features popular content which is representative of what a wide base of Internet users like the most. Content displayed in the front page often receives several million views per day, opening up opportunity for adversarial users and advertisers to game the system [6]. The tension between the finite space of the front page and proliferation of UGC, also calls for effective content discovery channels for users to find interesting content. Moreover, online content voting networks include a social network: users establish friend relationships. The social network can affect the way users share, search, browse and rate content; on the other hand, users' activities such as submitting and rating content can help build up friend relationships and thus boost their profile within the community [1].

To understand the nature and impact of online content voting networks, we in this paper analyze Digg, a popular social news aggregator and voting website. The main contribution of this paper is an extensive trace-driven analysis of users' digging activities on submitted stories. To this extent, we have collected a large amount of digg data spanning from the launch of Digg (2004/12/01) to 2009/04/16. We have also crawled the social network graph, more specifically, the large weakly connected component (WCC). To the best of our knowledge, this work is the first to extensively study user voting activities on content, and impact of the social network on content rating and vice versa. Our analysis yields very interesting findings regarding structural properties of the social network, the distribution of user diggs, content promotion (e.g., mark a story popular and surface it to the front page), content filtering, content censorship and content discovery.

The highlights of our work could be summarized as follows:

- We analyze the structural properties of Digg social network. We show that Digg differs from many other online social networks [18] in that it has low link symmetry and weak correlation of indegree and outdegree, and nodes tend to connect to other nodes with different degree from their own.

- By analyzing the correlation of users' social relationships and their digging activities, we show that users with more visibility (i.e., befriended by more users) within the community tend to vote more frequently. Users are not equally powerful and top users with high visibility tend to have high digging power in making stories popular.

- We reveal that the number of diggs and digg rate are influential to promotion of a story, and provide evidence of content censorship. Our findings also indicate existence of spam diggs.

- Two main information filters are present in Digg: the friends interface filters stories by the friends' activities and the story promotion algorithm sifts stories by users' aggregate votes. We show that information filtering impacts on users viewing and rating content. The promotion algorithm has substantially more influence on users than the friends interface.

- Leveraging our findings, we provide insight into design of the story promotion algorithm and a recommendation-assisted content discovery mechanism that helps users to find interesting content.

The remainder of the paper is organized as follows. Section 2 describes background on Digg and measurement methodology. We analyze structural properties of the social network in Section 3 and user digging activities in Section 4. Implications of our findings are discussed in Section 5. Section 6 provides related work and we conclude the paper in Section 7.

## 2. BACKGROUND AND MEASUREMENT METHODOLOGY

### 2.1 Background

Digg is a popular online news aggregation site where users submit and rate stories. When a user submits a story, the story is first placed on the *upcoming stories* section, which is the place where users browse recently submitted stories and digg what they like the best. If a story receives enough diggs (In the rest of the paper, we use *vote* and *digg* interchangeably) and meets promotion requirements, the story gets promoted and is moved to the *popular stories* section which we all the *front page*. In the front page, stories are more visible to the community and can receives several million visits per day.

Digg includes a social network: A registered user can invite other registered users as her friends; a user can also be befriended by other users who become her fans. The friend and fan links create a directional graph among users. Through the *friends* interface, Digg allows users to track friends' activities (stories they recently submitted or voted for).

To investigate structural properties of Digg social network, we focus on the large weakly connected component (WCC) as it is structurally the most interesting part of a social network [18]. To investigate nature and impact of user digging activities, we collect digg data of the users in the WCC. As will be shown later, the majority (about 90.75%) of diggs submitted by entire community come from the WCC. In subsequent sections, we describe the process of graph crawl and data collection, followed by high-level statistics of our crawled data.

### 2.2 Crawling the WCC

We used the Digg API "List Users" to crawl the WCC. Our crawl started with the user "kevinrose", the founder of Digg and inserted the user name into an initially empty queue. At each step, our crawl script removed a user from the queue, retrieved a list of the user's friends and fans, and added unvisited friends and fans into the queue. The crawl script continued until the queue is exhausted. Digg limits the rate at which a single IP address can download information. It took about one week to crawl the WCC and The crawling process ended on March 16, 2009.

### 2.3 Data Collection

Most of the data analyzed in this paper is about user diggs. Via the Digg API "List Events", for each user in the graph we fetched her diggs submitted between 2004/12/01 and 2009/03/16. We call this digg data trace *Primary Trace* (PT). In addition, we collected one month worth of digg data trace spanning from 2009/03/17 to 2009/04/16, which we call *Secondary Trace* (ST). ST is used to examine correlation of the user graph and user digging activities because: (1) the user graph was evolving as users joined over the time period of PT; and (2) the crawled graph is a snapshot of Digg social network around 2009/03/16 and is assumed to be relatively stable over the duration of ST.

### 2.4 High-level Statistics

Table 1 shows some high-level statistics of the data we collected. Other statistics of the data, such as diggs received by upcoming and popular stories in PT, will be presented in subsequent analyses. Note that diggs submitted by users in WCC constitute 90.75% of total diggs submitted by the entire community. This shows that WCC is not only structurally the most interesting component in the social network, but also logically the most valuable piece in analysis of user diggs. In the rest of the paper, we use *node* and *user* interchangeably.

**Table 1: High-level statistics of Digg crawl.**

| | |
|---|---|
| # of nodes in WCC | 580,228 |
| # of friend links in WCC | 6,757,789 |
| Avg. # of friends per user | 11.65 |
| Frac. of links symmetric | 39.4% |
| Duration of PT | 2004/12/01 - 2009/03/16 |
| # of diggs in PT | 154,129,256 |
| Avg. # of diggs per user in PT | 265 |
| Duration of ST | 2009/03/17 - 2009/04/16 |
| Frac. of users in WCC dugg in ST | 0.22 |
| # of submitted stories in ST | 257,536 |
| # of popular stories in ST | 4,571 |
| # of upcoming stories in ST | 252,965 |
| Frac. of diggs submitted by WCC | 90.75% |

## 3. ANALYSIS OF NETWORK STRUCTURE

In this section, we characterize the structural properties of the Digg social network, to answer a major question: Does the Digg social network show similar structural characteristics with other online social networks [17, 18].

### 3.1 Link Symmetry

Links in many previously studied online social networks are directed and therefore a user may link to any other users she wishes. For instance, a user may invite any other users to be her friends. Upon acceptance of the invitation, the invitee may reciprocate by pointing back to the inviter (often without scrutiny), thereby creating a high degree of link symmetry. Previous studies [18, 17] have observed a significant level of *reciprocity* in Flickr, LiveJournal [4], YouTube, Yahoo! 360 [7] and Okurt [5], ranging from 62% to 100%.

Table 1 shows reciprocity of the Digg graph is 39.4%, far lower than that of the aforementioned online social networks. The Web graph does not show a high level of link symmetry among web pages, and thus search engines leverage this to identify reputed source of information (pages with high indegree tend to be authorities) to rank search results (i.e., PageRank [20]). As will be discussed later, our findings suggest Digg possibly suffers Sybil attacks by which attackers create many identities to digg a story in order to surface it to the front page. The low level of link symmetry in Digg graph may allow us to defend against Sybil attacks by using PageRank-like algorithms to rank users and weigh their votes.

### 3.2 Power-law Node Degree

One striking property of online social networks is that their node degree distribution follows a power-law. That is, the majority of nodes have small degree while a few nodes have significantly higher degree. Figure 1 shows the indegree and outdegree complementary cu-mulative distribution functions (CCDF) for Digg social network. Digg exhibits behavior consistent with a power-law network. For the power-law distribution of node indegree, the most straightforward explanation is the *preferential attachment* process: the probability of a user $i$ connecting to a user $j$ is proportional to the number of $j$'s existing fans (or incoming links).

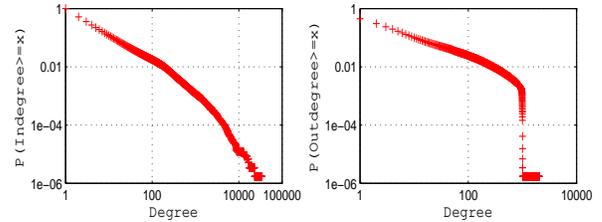

**Figure 1: Log-log plot of indegree (left) and outdegree (right) complementary cumulative distribution functions (CCDF). Digg network shows properties consistent with power-law networks.**

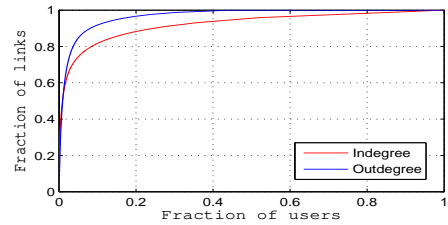

**Figure 2: Plot of the distribution of links across nodes.**

Compared to the previously studied online social networks [18], Digg however has a less number of high-outdegree nodes and their outdegree is significantly lower. Note a sharp drop at the degree of $1,025$. We have conjectured that Digg users not passionate in building an very large number of friend links is mainly because submitting and digging stories are probably more effective to boost their profiles within the Digg community [1].

Figure 2 plots the distribution of incoming and outgoing links across nodes in Digg graph. Digg shows similar distributions for incoming and outgoing links. For example, about 1.5% of users account for 60% of all incoming and outgoing links. The difference between the two curves mainly comes from the fact that a significant portion (55.21%) of users do not have friends.

### 3.3 Correlation of Indegree and Outdegree

In social networks such as YouTube, Flickr and Live-Journal, the nodes with high outdegree tend to have high indegree. For example, in all these three networks, the top 1% of nodes ordered by outdegree has a more than 65% overlap with the top 1% of nodes ranked by indegree [18]. The left plot in Figure 3 shows the ex-tent of the overlap between the top $x\%$ of nodes sorted by indegree and outdegree. Digg exhibits less overlap between the top $x\%$ of nodes ranked by indegree and outdegree. For instance, the top $1\%$ of nodes ordered by outdegree has a $58\%$ overlap with the top $1\%$ of nodes ranked by indegree. We have extrapolated that many high-indegree users have boosted their visibility to the community by submitting and digging stories, instead of aggressively making friends.

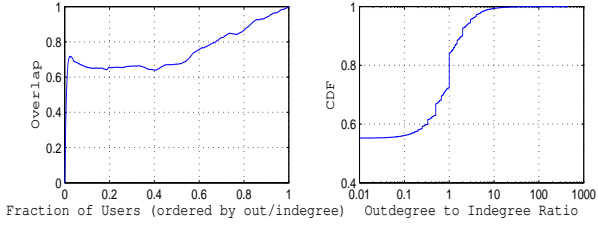

**Figure 3: Plot of the overlap between top $x\%$ of nodes ranked by outdegree and indegree (left) and CDF of outdegree to indegree ratio (right).**

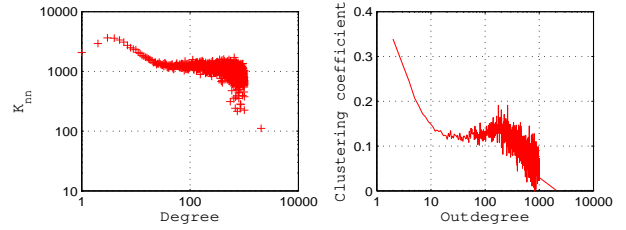

**Figure 4: Log-log plot of the outdegree versus the average indegree of friends (left) and average clustering coefficient of users with different outdegrees (right).**

In addition, we examined the indegree and outdegree of individual nodes in Digg. The right plot in Figure 3 shows the cumulative distribution of outdegree-to-indegree ratio for Digg. The CDF for Digg differs from those of YouTube, LiveJournal and Flickr in two ways: (1) about $55.21\%$ of nodes have outdegree of zero; and (2) about $14.56\%$ of nodes have an indegree within $20\%$ of their outdegree while the percentage for the other three social networks is more than $50\%$ [18].

In summary, Digg exhibits much weaker correlation of indegree and outdegree than the other three social networks. This can be explained by a much lower level of link symmetry in Digg social network.

### 3.4 Link Degree Correlation

Which users tend to connect to each other in Digg? To answer this question, we used the *joint degree distribution* (JDD) which is approximated by the degree correlation function $K_{nn}$. $K_{nn}$ denotes a mapping between outdegree and the average indegree of all nodes connected to nodes of that outdegree: an increasing $K_{nn}$ implies a tendency of higher-degree nodes to connect to other high-degree nodes while a decreasing $K_{nn}$ indicates the opposite trend. The left plot in Figure 4 depicts $K_{nn}$ for Digg. Unlike Flickr, LiveJournal and Okurt observed in [18], Digg exhibits the "celebrity"-driven nature. That is, there are a few extremely popular users in Digg to whom many unpopular users link.

To further explore this phenomenon, we calculated the assortativity coefficient $r$, a measure of the likelihood for nodes to connect to other nodes with similar degree. The assortativity coefficient $r$ value lies between $-1$ and $1$; a big $r$ indicates tendency of nodes to link to nodes of similar degree and a negative $r$ implies that nodes tend to link to nodes with very different degree from their own. We found that the assortativity coefficient $r$ for Digg is $-0.019$.

### 3.5 Clustering Coefficient

Next, we explore connection density of the neighborhood of a node, which is quantified by the *clustering coefficient*. Formally speaking, the clustering coefficient of a node with $N$ neighbors is defined as ratio of the number of directed links existing between the node's $N$ neighbors and the number of possible directed links that could exist between the node's neighbors ($N(N-1)$). The clustering coefficient of the Digg graph is the average of individual nodes' clustering coefficients, with a typical value of 0.218. This is consistent with those of YouTube, Okurt, Flickr and LiveJournal (ranging from 0.136 to 0.330) [18]. It shows that users in social networks tend to be introduced to other users via mutual friends, increasing the probability that two friends of a single user are also friends. The right plot in Figure 4 shows the clustering coefficients of nodes with respect to their outdegree. Nodes of low outdegree have higher clustering coefficients, indicating significant clustering among low-outdegree nodes. High-outdegree nodes, on the other hand, show much lower clustering coefficients due to their large number of diverse friends.

### 3.6 Summary

We conclude this section with a brief of summary of important structural properties of the Digg social network we observed in our data:

- The degree distributions follow a power law. Unlike many other social networks, Digg has a less number of high-outdegree nodes and their outdegree is significantly lower; this is explained by the fact that submitting and digging stories are more effective to boost their popularity within the community.

- Compared to many other social networks, Digg

shows a much lower level of link symmetry and weaker correlation of indegree and outdegree. Nodes tend to connect to nodes with very different degree from their own.

- Low-outdegree nodes have higher clustering coefficients than high-outdegree nodes. This hints that Digg seems to be composed of a large number of highly connected clusters consisting of low-degree nodes, and these clusters link to each other by a relatively small number of high-degree nodes.

# 4. ANALYSIS OF CONTENT VOTING ACTIVITIES

In Section 3 we have investigated structural properties of the Digg social network. In this section we focus on users' content voting activities. Unlike many other online social networks that are centered on building social relationships and sharing information, Digg, a news aggregator site, is centered around user voting on submitted stories to make them popular or not. Users vote on stories for different goals. Some users digg stories they like the best and share them with other users; some users digg stories as a means to boost their profiles within the community (e.g., become a friend of the users who submitted the stories they dugg); and some other users attempt to game Digg to make some articles (e.g., advertisement, and phishing articles) promoted to the front page to receive several million page views per day (often for profits). As a result, it is significantly important to examine users' voting activities which is not only helpful to understand user behavior and how stories get promoted, but also beneficial to system designs including story promotion algorithms and resource allocations (e.g., promotion of stories will attract more and more user attention including diggs and comments which need more resources to handle).

## 4.1 Statistics of User Diggs

First, we quantify the total number of diggs and daily average number of diggs submitted by individual users in PT; we also measure the total number of diggs submitted by individual users in ST. Figure 5 depicts three CCDFs of the total and daily average diggs [1] across users. The top and middle plots represent the total number of diggs in PT and ST respectively; the bottom one denotes the daily average number of diggs in PT.

Several important observations can be made: (1) The shapes of the two plots for total diggs closely match, implying stable digging activities across users. (2) A significant portion of users do not vote. For example, 20.2% and 78.0% of users did not vote in PT and ST, respectively. (3) A few users are very active in digging stories, e.g., a couple hundred diggs per day; while an overwhelming majority of users cast less than one vote per day.

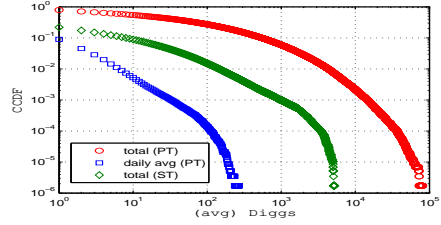

**Figure 5: Log-log plot of total diggs (top 2) and daily average diggs complementary cumulative distribution functions (CCDF).**

For the overwhelming majority of users, since they digg infrequently, most of them probably follow a "read-then-digg" pattern: Read an article first and then digg it if they like it. For those users passionate in digging stories, however, they may follow a different pattern. We illustrate this by an example. Suppose an active user diggs 200 stories per day and follows a "read-then-digg" pattern. Also assume that reading an article (probably also spending time finding the stories) takes 2.5 minutes. Then, the active user has to spend more than 8 hours a day. This is a large amount of time invested in Digg unless we suspect the user of either not following the "read-then-digg" pattern (e.g., digging stories solely to boost her profile in the Digg community) or using an automatic script.

Thus, we analyze the time intervals between a user's consecutive diggs. The left plot in Figure 6 displays PDF of time intervals between consecutive diggs by users in PT. Note that the time intervals follow a power-law distribution. Most of diggs by individual users are submitted close in time. In particular, about 35% of diggs are submitted within one minute or less [2] after their previous diggs. Such diggs unlikely result from the "read-then-digg" pattern because of the time to (find and) read an article.

Diggs submitted by automatic scripts are most likely to be close in time. Golder et al. have used 5 seconds as the threshold of inter-message time for automatically generated messages and manual ones [14]. They classify those Facebook messages written within 5 seconds of the previous messages as spam. We do not intend to argue whether or not 5 seconds are an appropriate threshold to mark user diggs as spam or to infer user diggs generated by automatic scripts. The right plot in

---

[1] The average diggs for a user is based on the time period from the first time the user dugg to the end time of the trace. We believe it is more reasonable to begin with the first time the user dugg than the user's registration time.

[2] If the inter-digg time interval is less than 1 minute, we round it up to 1 minute.

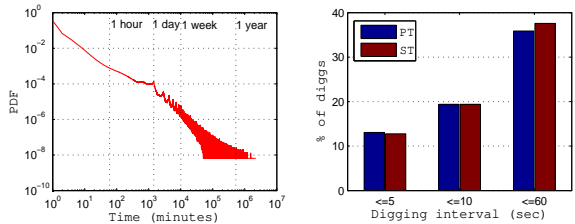

**Figure 6: Log-log plot of PDF of time intervals between users' consecutive diggs in PT (left) and percentages of diggs submitted within 5, 10 and 60 seconds of their previous diggs (right).**

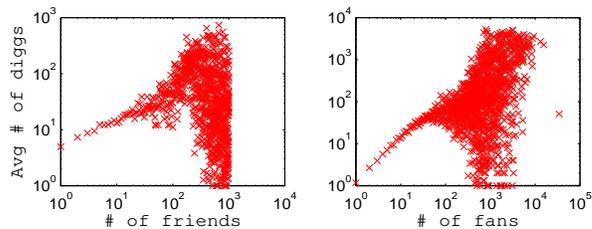

**Figure 8: Avg. number of diggs versus number of social links. The left plot is for number of friends and the right plot is for number of fans. The trace ST is used.**

Figure 6 shows percentages of diggs submitted within 5, 10 and 60 seconds of inter-digg time. Both PT and ST have similar characteristics: A significant portion of diggs are indeed submitted close in time; for example, over 12.75% of diggs are generated within 5 seconds of their previous diggs.

Inter-digg time interval is important to characterize a user's digging behavior. A user whose diggs are very close in time, very likely diggs a story not upon its content but for other purposes. For example, to boost her profile in the community, a user diggs stories hoping story submitters to reciprocate by connecting to her; a user may follow her friends to digg what they have dugg. Also, people are gaming Digg to get their stories (e.g., advertisement) into the front page. *Subvert and Profit* [6] is a service to sell home page placement on Digg by charging advertisers for a vote; users may sell their votes to such service companies for profits, thereby creating spam diggs.

User-driven social content websites like Digg, should have to identify and handle spam votes if they want to retain democracy of content rating which is one of the driving factors contributing to success of today's user-driven social content websites. We conjecture that user inter-digg time interval may be used to detect spammers and digg spam, therefore making story promotion algorithms resilient to spam diggs.

### 4.2 Correlation of Users' Diggs and Social Links

Next, we provide data to answer two questions: (1) Do people digg more actively if they have more friends? (2) Do people digg more actively if they are befriended by many others? In Figure 7 we depict the average number of diggs (the left two plots) and the average number of daily diggs (the right two plots) against the number of friends and fans per user respectively. We see that the number of friends does influence the users with up to 200 friends. That is, people digg more as the number of friends increases until reaching 200 friends. Beyond 200 friends, there is no strong correlation of the num-ber of friends and digging activity. However, we observe strong correlation of the number of fans and digging activity: People digg more with increasing number of fans. This can be explained by three factors: (1) People increase their visibility to the community through more diggs, and thus attract more users to connect to them and to become their fans; (2) People with more fans respond to "celebrity" pressure by digging more; and (3) People with more fans are likely to have been in the network longer, thereby accumulating more diggs. The first two factors are like the "chicken-and-egg" problem and it is hard to tell which dominates. To investigate the last factor, i.e., impact of user age (the time since registration) on the correlation of the number of fans and digging activity, Figure 8 shows the average number of diggs vs number of friends and number of fans in ST. ST minimizes noise of the last factor in the correlation. Similar characteristics are observed, though the correlation of the number of fans and digging activity is slightly diluted.

### 4.3 Story Promotion vs User Diggs

The lifecycle of stories in Digg is as follows. A newly submitted story first goes to the *upcoming stories* section, displayed in reverse chronological order of submission time. If the story accumulates enough votes shortly and meets promotion requirements, it will be marked as a *popular story* and promoted to the front page, thereby becoming more visible to the community and receiving more visits. Otherwise, the story will be pushed down in the upcoming stories section as it ages, thereby becoming less visible and finally getting "buried".

While unrevealed to the public, the promotion algorithm is crucial to Digg. On one hand, it should promote the content that a wide base of users like the most; On the other hand, it should be resilient to gaming [6] and Sybil attacks. For example, in ST we have found some upcoming stories, a significant portion of whose received diggs come from new voters who registered on the same day of the story's submission time. In the crawled Digg graph, we have also found many users whose usernames only differ in their trailing dig-its and whose registration times are close in time. We do not intend to use these findings as evidence of Sybil attacks. But, the promotion algorithm design should take attacks into account because the most profitable avenue for the attacks lies in the fact that the promoted content receives several million visits per day.

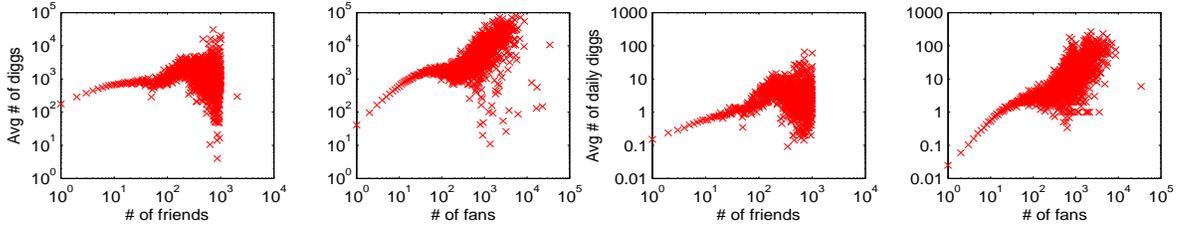

**Figure 7: Avg. number of (daily) diggs versus number of social links. The left two plots are for avg number of diggs vs number of friends, and avg number of diggs vs number of fans; the right two plots are for avg number of daily diggs vs number of friends, and avg number of daily diggs vs number of fans. The trace PT is used.**

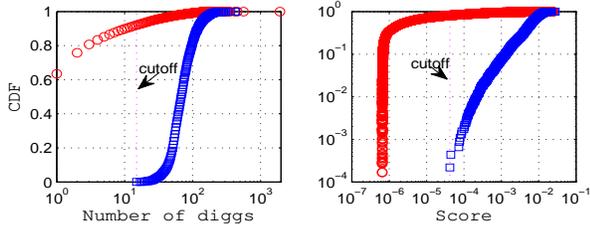

**Figure 9: CDF of number of received diggs (left) and CDF of scores (right) for upcoming ("circle") and popular ("square") stories in ST. The number of diggs for an upcoming story includes all diggs received since the submission of the story until the end time of ST; The number of diggs for a popular story includes only the diggs received prior to its promotion time.**

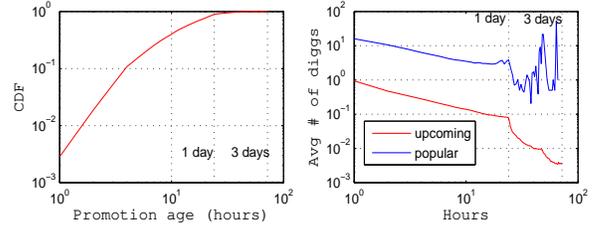

**Figure 10: Left: CDF of promotion ages across popular stories in ST. Right: Average digg rates for upcoming and popular stories in ST. Each data point represents average dig rate for upcoming or popular stories at each age (by hours).**

To this end, we use ST to examine impact of user diggs on story promotion [3]. Specifically, we have identified that *number of diggs* and *digg rate* play influential roles in story promotion.

The left plot in Figure 9 shows CDF of number of diggs for upcoming and popular stories in ST. Apparently, the number of diggs strongly influences whether a story can become popular or not. Note that 92.1% of upcoming stories received less than the cutoff value of 15 diggs while all popular stories received 15 diggs or more before promotion. Surprisingly, 7.9% of upcoming stories received same number of diggs as popular stories or even more (e.g., one upcoming story received 1,936 diggs) but did not get promoted. We conclude that the number of diggs is not the only factor contributing to story promotion. Note that for popular stories we consider only the diggs received before their promotion while for upcoming stories we count all the diggs received over the duration of ST. We have extrapolated that *digg rate*, defined as the number of diggs received by a story for each hour, affects promotion.

The left plot in Figure 10 shows CDF of promotion ages across popular stories. We can see that most of stories (88.4%) get their promotion at age of one day or younger. All stories become popular within 3 days after submission. We conclude that if a story will become popular, it will get promotion very soon. This suggests that digg rate is influential to promotion.

The right plot in Figure 10 shows average digg rates at each age (in terms of hours) for upcoming and popular stories. It only plots data up to 3 days since all popular stories get promotion by age of 3 days. For a popular story, if it gets promoted at age of $m$ hours, then its diggs received after promotion are not counted and the story is excluded from calculating the data points after $m$ hours. Simply put, we only show the digg rate *prior to promotion* for popular stories. From the figure, we can see that popular stories before promotion, receive a digg rate which is one order of magnitude higher than that of upcoming stories. We believe that the digg rate, particularly in the initial several hours, is very important to story promotion. In other words, if a story does not get sufficient diggs at its early age, it will then lose visibility and finally get buried.

We suspect that Digg's promotion algorithm treats each individual vote equally. Otherwise, the promotion algorithm is susceptible to gaming and Sybil attacks as attackers can create many identities to vote for a story in order to advance it to the front page. As discussed in Section 3.1, Digg shows only 39.4% link symmetry far lower than that of other online social networks (62% - 100%). Low level of link reciprocity makes it not only difficult for attackers to foster trust (friend) links from others, but also easy to identify reputed source. Inspired by PageRank [20] to measure importance of web pages, we used PageRank algorithm to weigh individual diggs. First, we ran PageRank in the crawled graph and computed a PageRank value $w_i$ for each user $i$. Then, $w_i$ is used to weigh user $i$'s diggs. For an article $A$ with $m$ votes, the score is calculated by $score(A) = \sum_{k=1}^{m} w_k$. The right plot in Figure 9 shows CDF of scores for upcoming and popular stories and it shows similar characteristics with the left plot in Figure 9. One striking difference is that those upcoming stories which received more diggs than any of popular stories are subsumed by the score curve of popular stories. This indicates that many diggs on those upcoming stories come from insignificant users. Again, the digg rate adversely affects promotion of some upcoming stories with scores comparable to those of popular stories. In the subsequent section, we will show other promotion-related factors.

## 4.4 Controlled Democracy?

It is widely believed that the content seen in the front page is representative of what a wide base of Digg users like the most, which leads us to believe that Digg is a "democratic" community: the voice of the majority is heard.

However, after examining a sample of upcoming stories whose received diggs and digg rates are comparable to or even more than those of popular stories, we have found evidence of content censorship on advertisement, phishing articles and articles "offensive" to Digg. Figure 11 plots the number of diggs accumulated at different ages for upcoming and popular stories as well as a censored story. Despite that the diggs and digg rate of the censored story at various ages are nearly one order of magnitude higher than those of popular stories, it is still not promoted at the time of writing (nearly two months after story submission). The story will be unlikely promoted noticing the flat tail of its curve. Our findings indicate that Digg may bury stories by censoring content before they are promoted to the front page.

Next, we examine ST to answer a question: Are Digg users equally powerful or a few are more influential? For a user $i$, let $P_i$, $P'_i$, and $U_i$ denote the number of $i$'s diggs on popular stories prior to their promotion (i.e., user $i$'s diggs contribute to the promotion of the popular stories), the number of $i$'s diggs on popular stories after their promotion, and the number of $i$'s diggs on upcoming stories, respectively. Then, we use *digg success rate*, defined as $\frac{P_i}{P_i + U_i}$, to measure user $i$'s digging power (i.e., capability of making an upcoming story popular). Note that $P'_i$ is disregarded in the digg success rate.

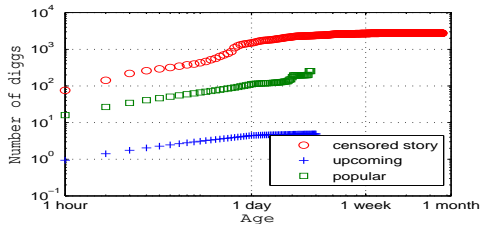

**Figure 11: The *average* number of diggs received at different ages for upcoming and popular stories as well as the number of diggs received for a censored story in ST.** The number of diggs for an upcoming story includes all diggs received from the submission of the story to the age of 3 days; The number of diggs for a popular story includes only the diggs received prior to its promotion time. For the censored story, we include all diggs it received from submission to the end time of ST.

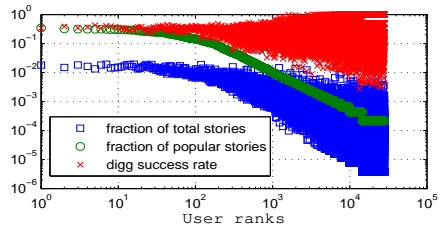

**Figure 12: Digging behavior of users ranked by $P_i$.** Let $P$ and $U$ denote the number of popular stories and upcoming stories in ST, respectively. For a user $i$, the *fraction of popular stories* is defined as $\frac{P_i}{P}$; The *fraction of total stories* is defined as $\frac{P_i + P'_i + U_i}{P + U}$, indicating the user's digg frequency.

Figure 12 shows digging behavior of the users who dugg in ST. Only 22.5% of the users in the Digg graph dugg on stories as shown in Table 1. The $x$-axis denotes user ranks ordered decreasingly by their $P_i$. It is straightforward that the *fraction of popular stories* drops as the user rank decreases. The top 100 users each contributes to the promotion of a significant portion (14.4−34.6%) of the popular stories by their diggs.

Two important observations can be made in this fig-ure. First, the top ranking users also have high *fraction of total stories*, which means they are among the most active users in digging. Second, the top 100 users all have good digging power with digg success rate ranging from 0.24 to 0.45. On the other hand, the low ranking users are very divergent in digging frequency and digging power: some users dugg infrequently but with very high success rate; some others dugg as much as some top users but with very low success rate. Overall, the top 100 users are more powerful because they not only digg the most but also promote stories the most. Further examination reveals that 76 of the top 100 users are among top 0.1% (correspondingly 89 users are among top 0.5%) of the total users ranked by PageRank algorithm in the crawled graph. Because high-indegree users tend to have high PageRank value and also they tend to digg more as shown in Section 4.2, the high digging power of these top 100 users result from their high visibility to the community. It is worth pointing out that 30.2% of the top 0.1% users (ranked by PageRank value) did not digg in the trace. This further renders the top 100 users (shown in the figure) more influential in story promotion. Interviews with Digg users [15] reveal that the top users might exploit their digging power to game the system.

## 4.5 Content Filtering vs User Diggs

A Digg user browses content in three ways: (1) using the *front* page to browse recently promoted stories; (2) using the *upcoming* page to view recently submitted stories; and (3) using the *friends* interface to see the stories her friends have recently submitted or voted for. The front page and the friends interface are essentially content filters, which sift content by the story promotion algorithm and *friends' taste* respectively. Undoubtedly, content filtering influences how people view and rate content.

The first question we raise is: Do friends' diggs influence one's votes? To answer the question, leveraging VSM [11] in IR algorithms we introduce a concept of *vote similarity* to quantify the influence of friends. For a user $u$ and her $m$ friends, let $X$ and $Y$ denote the user $u$'s and her friends' *digg vector* respectively. The digg vector $X$ (or $Y$) consists of $m$ (or $n$) $(s, f(s))$ pairs, where $s$ is the story the user (or her friends) dugg and $f(s)$ is the number of diggs on the story $s$ by the user (or her friends [4]).

For each pair $(s_i, f(s_i))$ in a digg vector, we replace it with $(s_i, w_i)$ by using the *dampened scheme* [21] where $w_i = 1 + \log f(s_i)$. Considering digg vector $X$, we normalize it as follows: For each $< s_i, w_i > \in X$, we normalize its weight $w_i$ and substitute it with a normalized weight $w'_i = \frac{w_i}{\sqrt{\sum_{j=1}^{m} w_j^2}}$.

---

[9] [4]If $k$ friends vote for $s$, then $f(s) = k$.

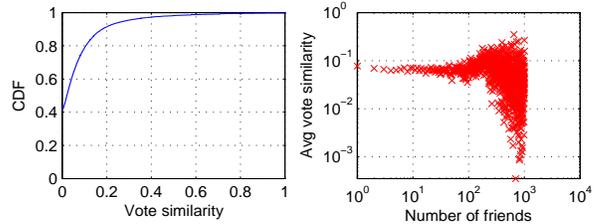

**Figure 13: The left figure plots CDF of vote similarity across users in ST. The right figure plots average vote similarity across the users with the same number of friends.**

Following the above process, we get normalized digg vectors $X'$ and $Y'$ for the user and her friends respectively. Then, the vote similarity between the user $u$ and her friends is:

$$Sim(X', Y') = \sum_{s_i \in X' \cap Y'} w'_{X',i} \cdot w'_{Y',i} \qquad (1)$$

The vote similarity lies in between $[0, 1]$: A high value means a user has dugg many same stories with her friends; a low value indicates the user and her friends have dugg very few stories in common. Note that the vote similarity exaggerates influence of friends because a user digging on a same story with her friends does not necessary suggest that the user follows her friends (e.g., a user dugg a same story with her friends in coincidence or a user dugg a same story earlier than her friends did). It represents the upper bound of friends' influence.

Figure 13 shows vote similarity of users with their friends. Two main observations can be made: (1) Most of (97.2%) users have vote similarity of less than 0.4 with their friends while a few (1.9%) users have vote similarity of more than 0.5 with their friends; (2) The vote similarity for users with fewer number of friends ($\leq$ 200) is consistently low while that for users with higher number of friends is divergent. Further examination reveals that this is consistent with the correlation of users' diggs and their number of friends in Figure 7: The number of diggs submitted by users with higher number of friends spans a wide range, thus resulting in a wide range of vote similarity. In summary, the friends interface does not have substantial influence on most users' digging behavior.

Next, we explore how the front page impacts users on viewing and digging stories. In addition to collected diggs in ST, we have also gathered the users' comments on the stories over the duration of ST, for better characterizing the users' activities on content viewing and rating. To quantify the impact of promotion on popular stories, it is straightforward: For a popular story which is promoted at age of $t$, we compare the visits (diggs and comments) received prior to promotion and those received from age $t$ to $2t$. For upcoming stories, we use $t = 72$ hours. This is because all popular stories are promoted within 72 hours as mentioned in Section 4.3. However, we do not claim that 72 hours are the choice of the most appropriate. Table 2 shows impact of the front page on content viewing and rating in terms of percentage increase. The front page placement significant boosts the number of votes and comments on the popular stories by $4.6 - 5.6$ times; the upcoming stories, sifted out by the promotion algorithm, rapidly lose 95% of user interactions. We thus conclude that story promotion, as an information filtering technique, significantly affects how people browse and rate information.

**Table 2: Impact of the front page on content viewing and rating.**

|                 | Upcoming | Popular |
|-----------------|----------|---------|
| Diggs           | $-95.2\%$ | $455.9\%$ |
| Comments        | $-94.1\%$ | $559.8\%$ |
| Diggs + Comments | $-95.1\%$ | $462.2\%$ |

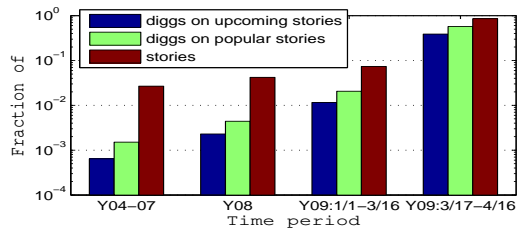

**Figure 14: Digg (story) vs submission time distributions of the stories dugg by users from 2009/03/17 to 2009/04/16.**

Figure 14 shows both digg and story distributions vs submission time distribution of the stories dugg by users from 2009/03/17 to 2009/04/16. The *diggs on upcoming stories* represents diggs on stories (including *prospective* popular stories prior to their promotion) while the *diggs on popular stories* represents diggs on stories after their promotion. The $y$-axis is in log-scale. Note that users dugg a small portion of very old stories (even 5 years ago), including upcoming and popular stories [5]. For the old popular stories (due to limited space in Digg's front page) and the old upcoming stories, they both are significantly less viewed and rated and thus get filtered. We have conjectured that users dugg them up via the friends interface. The small portion of the old stories and their diggs, confirms that influence of the friends interface is not substantial. Story age adversely affects diggs. For instance, recent stories (submitted from 2009/03/17 to 2009/04/16) absorb 96.1% of all users' diggs. Moreover, it confirms that popular stories draw significantly more votes than their counterparts thanks to content filtering by the story promotion algorithm, considering that the number of popular stories is far less than that of upcoming stories.

### 4.6 Summary

We end this section with a brief summary of our findings:

- An overwhelming majority of users digg stories infrequently while a small number of users are very active (i.e., a couple hundred diggs per day). Ex-amination of time intervals between each individual user's consecutive diggs suggests existence of digg spam and possible system gaming.

---

<sup>9</sup><sub>5</sub>Even the first submitted story in 2004/12/01 was dugg by the users.

- Users with more visibility (i.e., more fans) tend to digg more. Users are also more active in digging with increasing number of friends up to 200 friends; Beyond that, there is no strong correlation of the number of friends and digging frequency. Users are not equally powerful in digging stories. A few top users have high digging power largely due to their high visibility to the community.

- The number of diggs and digg rate are among determinants of story promotion. Evidence indicates that Digg is censoring content.

- Information filtering presented in Digg includes the friends interface and the story promotion. The friends interface influences users on viewing and rating content, but not substantially; the story promotion is substantially influential to users.

## 5. DISCUSSION

In this section, we discuss some implications of our findings.

**Design of Story Promotion Algorithm**. The story promotion algorithm is an important key to success of online content voting networks like Digg because it embodies a sense of *democracy*: It is Internet users' own judgment and passionate participation that make a story popular and surface it to the front page to receive several millions views per day. In the meanwhile, the story promotion algorithm is becoming a point of attack and gaming because the most profitable avenue for attacks and system gaming is to get one's content (e.g., advertisement) promoted for catching such a large amount of visits. For some reason Digg does not unveil the story promotion algorithm. While showing importance of the number of diggs and digg rate to story promotion, we believe that they are not all the pieces. A good promotion algorithm should not treat all the votes equally because of digg spam and Sybil attacks. A plausible solution is to weigh votes by evaluating credibility of voters. That is, we favor votes from trusted source. So, promotion of a story is not just simply summing up all the votes but weighted aggregate of votes.

Our findings have some interesting implications for gauging credibility of voters. Digg shows a much lower level of link symmetry than many other social networks. We may leverage the low level of link symmetry in Digg graph to identify reputed sources of votes, as search engines exploit this property to identify reputed Web pages [16]. We ran PageRank algorithm on the Digg graph and confirmed that high-indegree nodes tend to have high PageRank value [6]. Because Digg users do not tend to reciprocate their fans casually, exploiting the property of low link symmetry can defend against Sybil attacks: Attackers may be able to create as many identities as possible in a short time but it is difficult for them to build up many incoming links quickly, so they have low reputation scores and their votes are lightly weighted; Even a large amount of such votes will not subvert the promotion algorithm. However, well-recognized users (with a high number of fan links) can game the system as our findings reveal that these users are usually top users with high digg success rate. To counteract gaming by these top users, the promotion algorithm may take into account a user's voting history and dynamically adjust her reputation score (Additive Increase/Multiplicative Decrease as an example policy). As discussed earlier, inter-digg time can be used to detect spam diggs and content censoring can be used to identify those voters who have dugg on inappropriate content. When a user's spam diggs or inappropriate diggs reach a threshold, her reputation score is decreased multiplicatively. A similar policy can also be applied to story submission. As a result, the influence of a top user's votes will be suppressed upon misbehavior, making the promotion algorithm resilient to gaming by the top users.

**Recommendation-assisted Content Discovery.** One important mechanism in online content voting networks is content discovery which helps people find good content. People can find content in the front page for popular stories and upcoming page for recently submitted stories; however, the limited space of the front page and upcoming page does not meet all users' needs. The friends interface allows a user to find the stories her friends submitted and dugg, but our findings indicates that it does not substantially influence users on browsing and rating content. One plausible explanation is that users do not find interesting what their friends submitted and dugg. Indeed, most of users do not have much vote similarity with their friends as shown earlier. Thus, the friend list alone cannot provide good recommendation on content. We can cluster system-wide users based on their digg and submission activities [7] (e.g., by using bottom-up hierarchical clustering), and like-minded voters are grouped into same clusters. Users can go to a recommendation page consisting of stories voted and submitted by their cluster members to discover interesting content.

# 6. RELATED WORK

The most relevant work is research on social networks which generally falls into two areas: some researchers have focused on examining graph theoretic properties of social networks [9, 17, 10, 18] and others have investigated usage patterns in social networks [12, 14, 19].

Adamic et al. [9] examine an early online social network, Club Nexus, at Stanford University and show that the network exhibits both small-world behavior and strong local clustering. Kumar et al. [17] analyze two online social networks Flickr and Yahoo! 360 and find that both contain a large strongly connected component. They also show a high level of link symmetry in the two networks, with $70.2 - 80\%$. Backstrom et al. [10] study group formation and evolution in Live-Journal and present models for group evolution. Mislove et al. [18] present a large-scale measurement study of structural properties in four online social networks: Flickr, YouTube, LiveJournal and Okurt and confirm the power-law, small-world, and scale-free properties of online social networks. They also reveal high level of link symmetry in the four network, with $62 - 100\%$. Our results show that Digg social network, while showing some similar structural properties with these online social networks, exhibits a much lower degree of link symmetry ($39.4\%$).

Cha et al. [12] present an extensive data-driven analysis on the video distribution, video popularity evolution, and content duplication in YouTube. Golder et al. [14] analyze message activities of online social networks of college students in Facebook, and reveal that messaging within Facebook exhibits robust and consistent temporal rhythms across campuses and across seasons. In recent work, Nazir et al. [19] study usage characteristics of three Facebook social network based applications, and highlight that a small fraction of users account for the majority of the activity within each application and a small number of applications account for the majority users on Facebook.

Chun et al. [13] investigate an activity network constructed from the comments users wrote in other guest-books and find that the structural properties of the activity network are similar to that of the social network in Cyworld. To address the Sybil attacks on content rating, Tran et al. [22] propose adaptive vote flow ag-gregation to aggregate users' votes and limit the number of bogus votes cast by attackers in vote aggregation no more than the number of attack edges.

---

[9] [6]The founder of Digg, Kevin Rose, had the highest number $(34, 217)$ of incoming links at the time of our crawl, and thus had the highest value of 0.011.

[9] [7]Submission of a story can be viewed as a vote on the story.

## 7. CONCLUSIONS

In this work we investigated the structural properties and digging activities of Digg, an online content voting network. The graph we crawled is WCC of the user graph and the data of digging activities we collected spans from 2004/12/01 to 2009/04/16. Our data shows that the Digg social network differs from many other online social networks in that it has a low level of link symmetry and weak correlation of indegree and outdegree, and nodes tend to connect to nodes with different degree from their own. We reveal that the number of diggs and digg rate play an influential role in story promotion, and we also present evidence of content censorship in Digg. Our findings suggest that the story promotion algorithm, as an information filter, significantly influences on viewing and rating content while the friends interface is not substantially influential to users. We show that users with high visibility tend to digg stories more actively and that top users are those who have high visibility and digg power. Our data also suggests existence of spam diggs. We have outlined how these findings may affect the story promotion algorithm and content discovery for online content social networks.

In our next step, we plan to investigate the properties of the activity network constructed based on user interactions: a directional edge from $A$ to $B$ is formed if user $A$ dugg or wrote a comment on $B$'s stories; and the edge weight is proportional to the number of diggs and comments $A$ generated for $B$'s stories. The activity network may lend insight into the actual dynamics of interaction between users in online content rating networks.

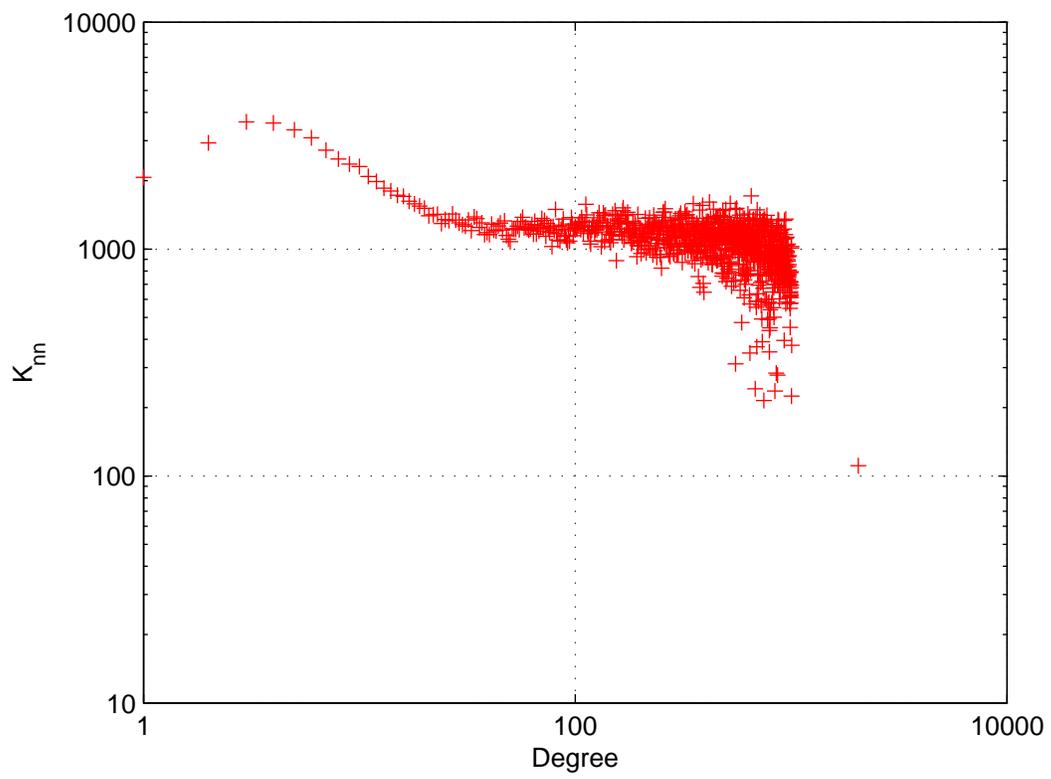

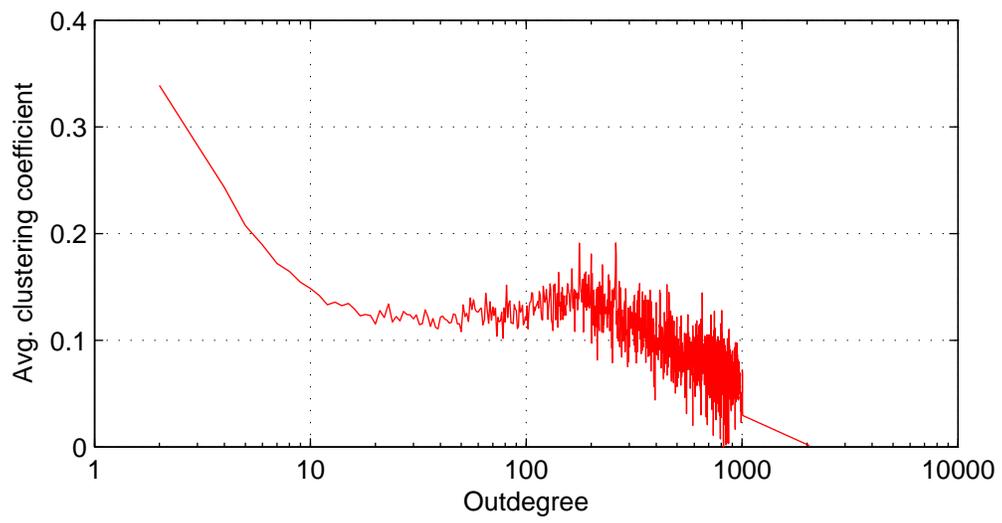

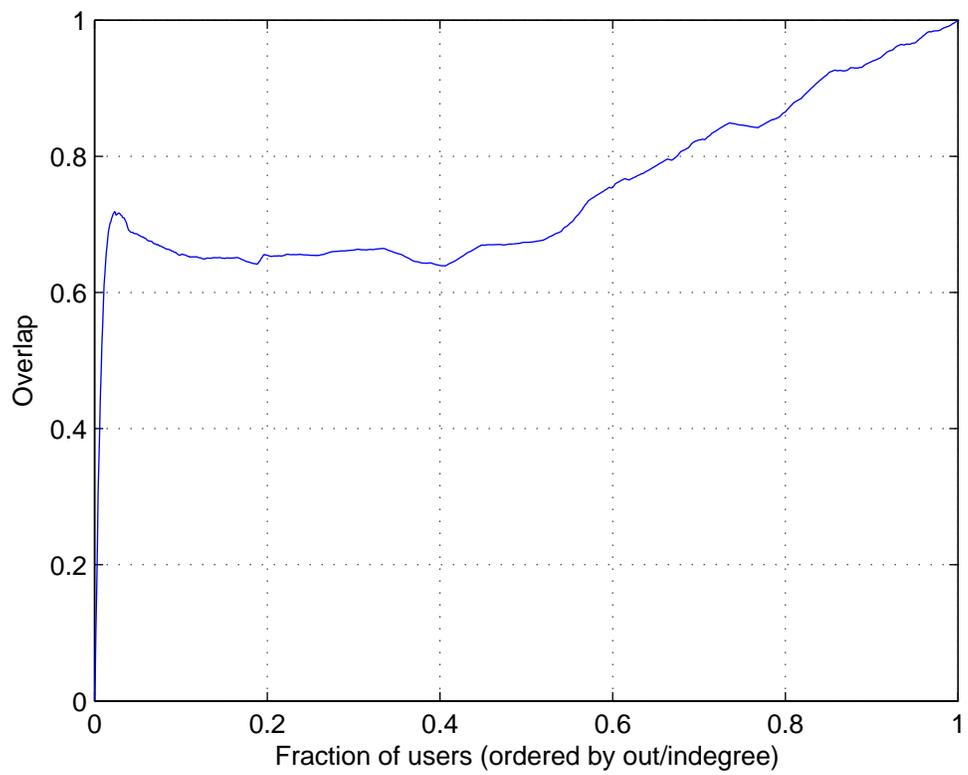

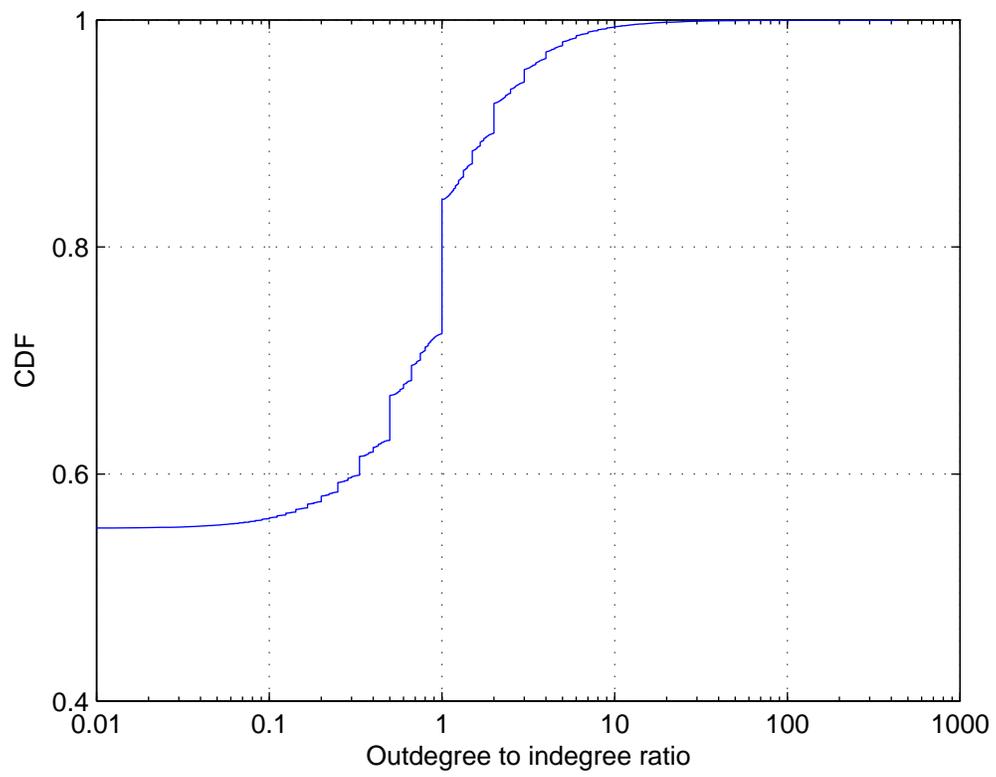

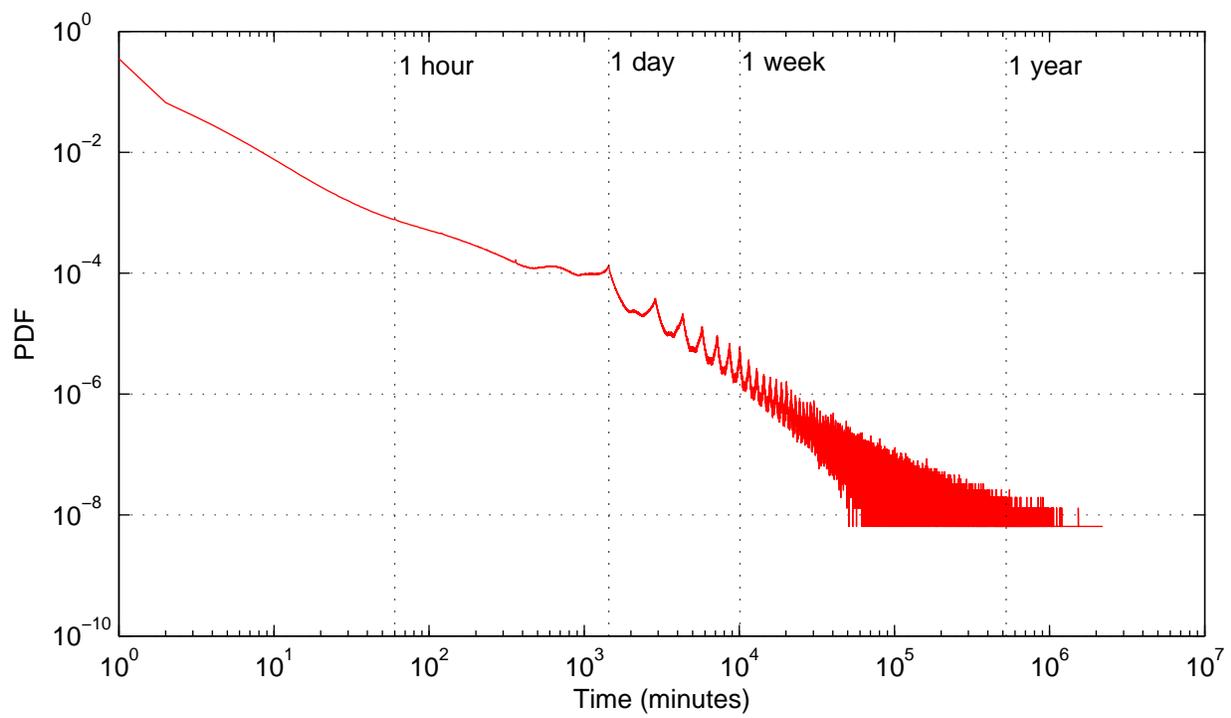

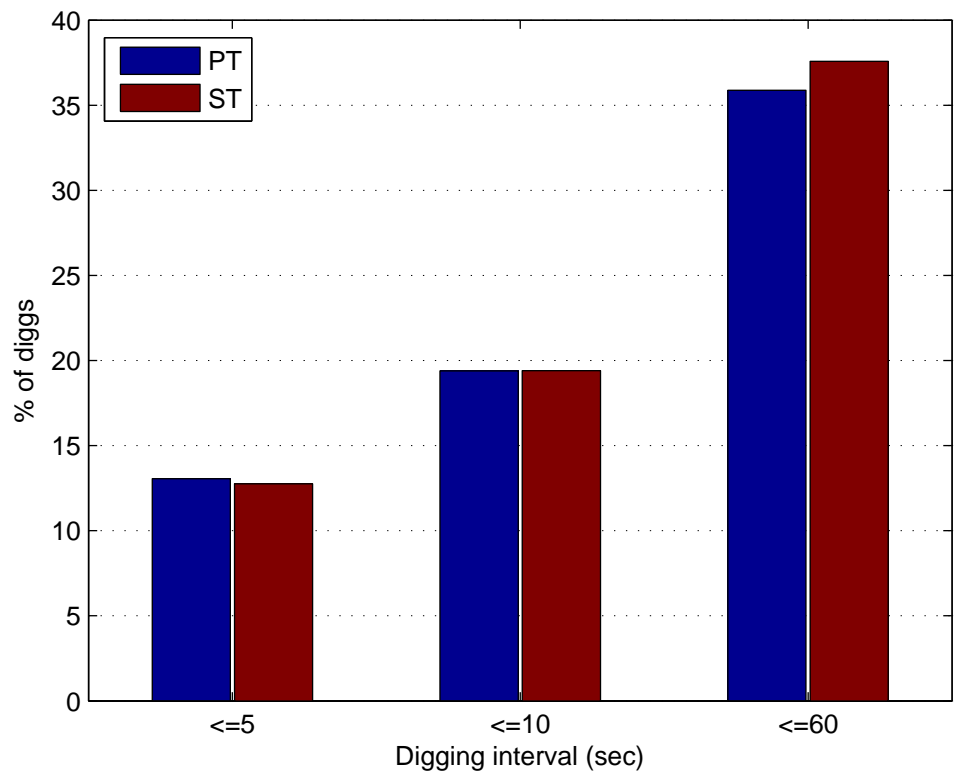

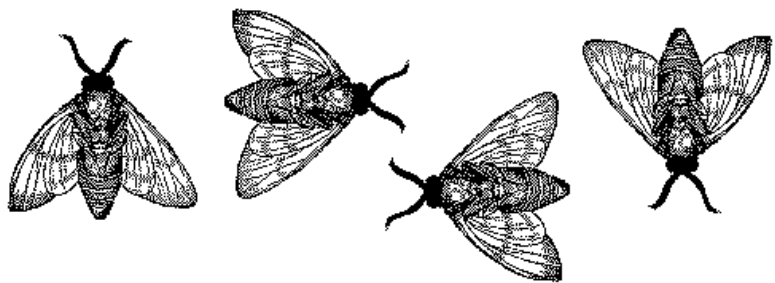

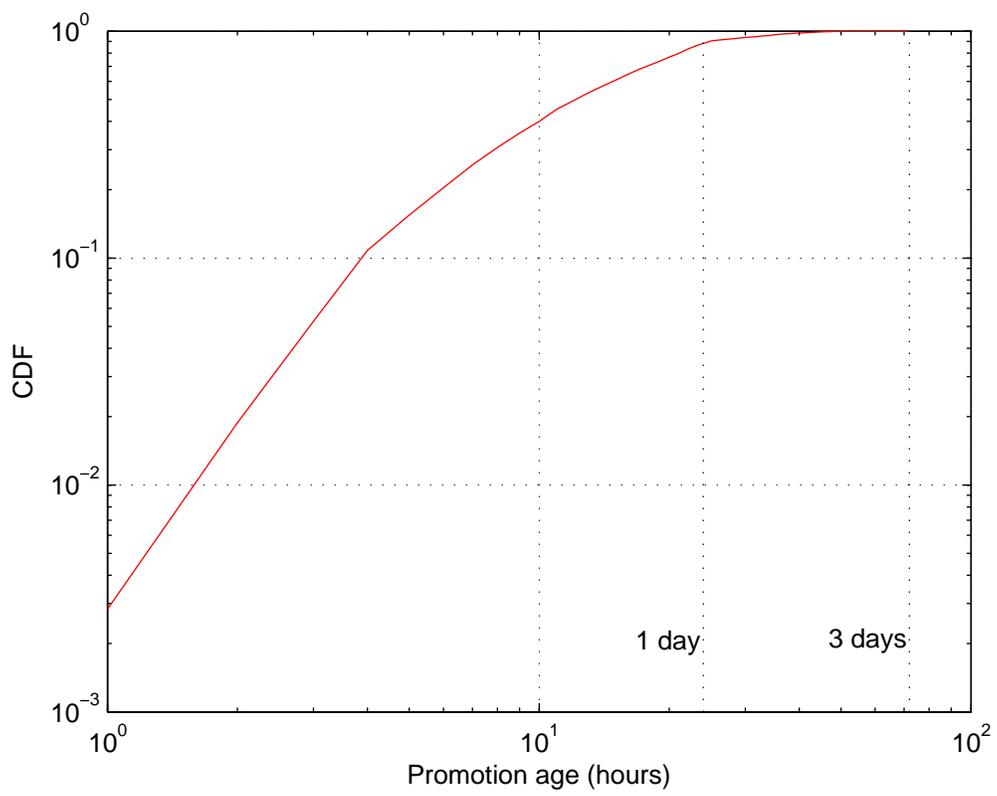

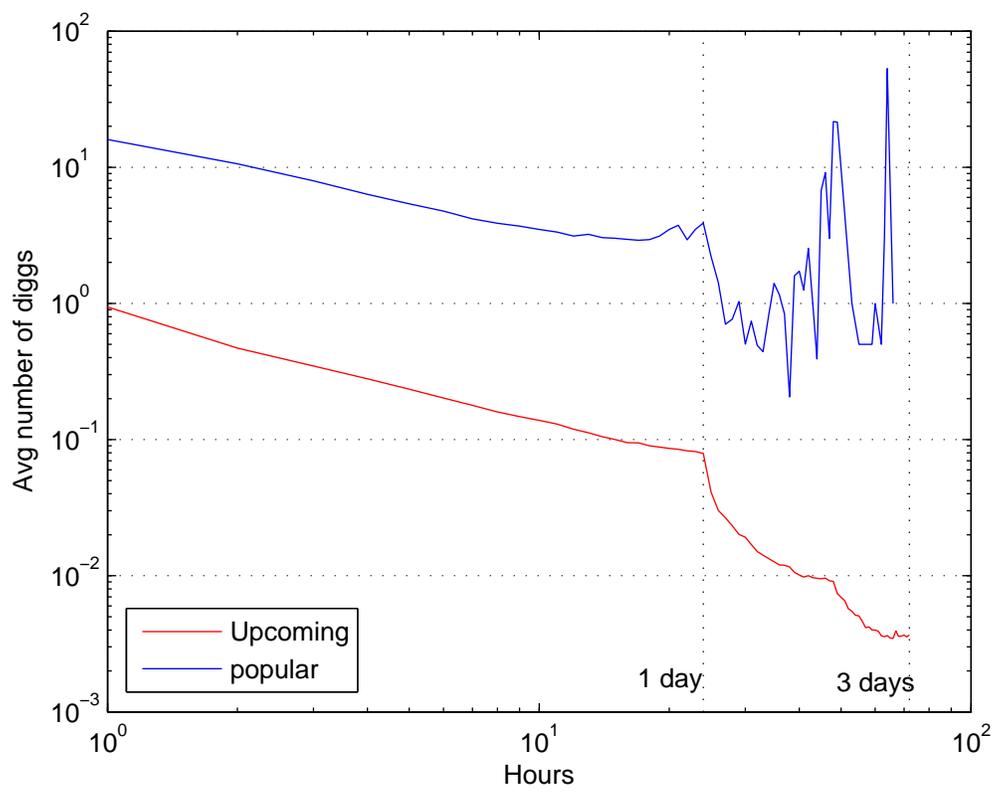

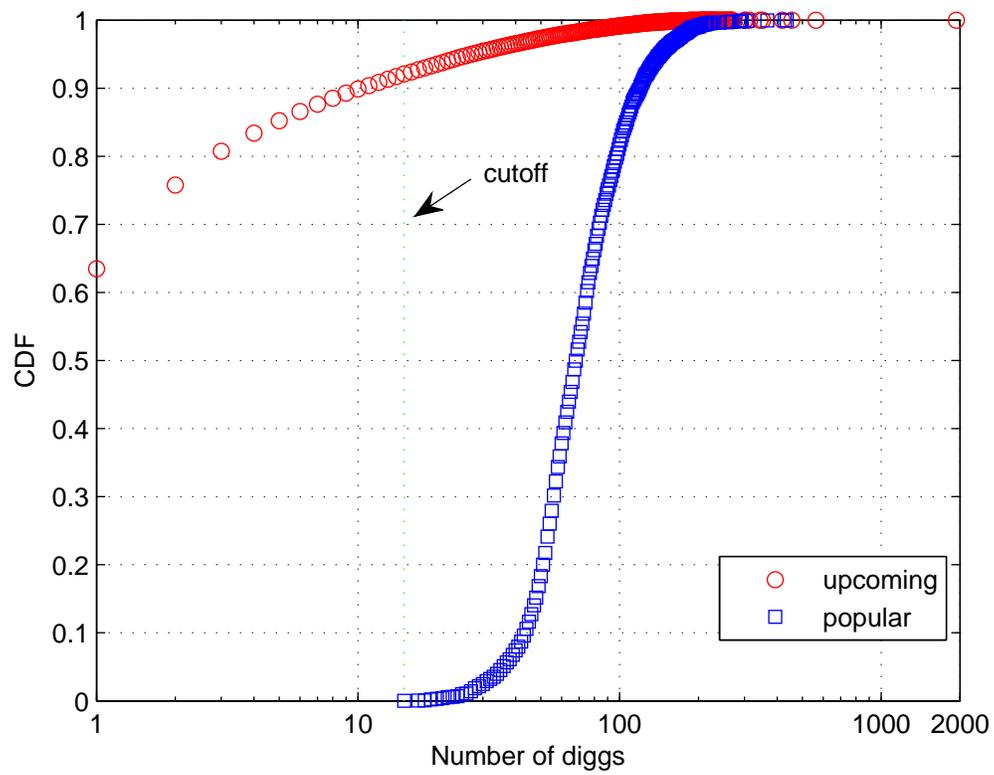

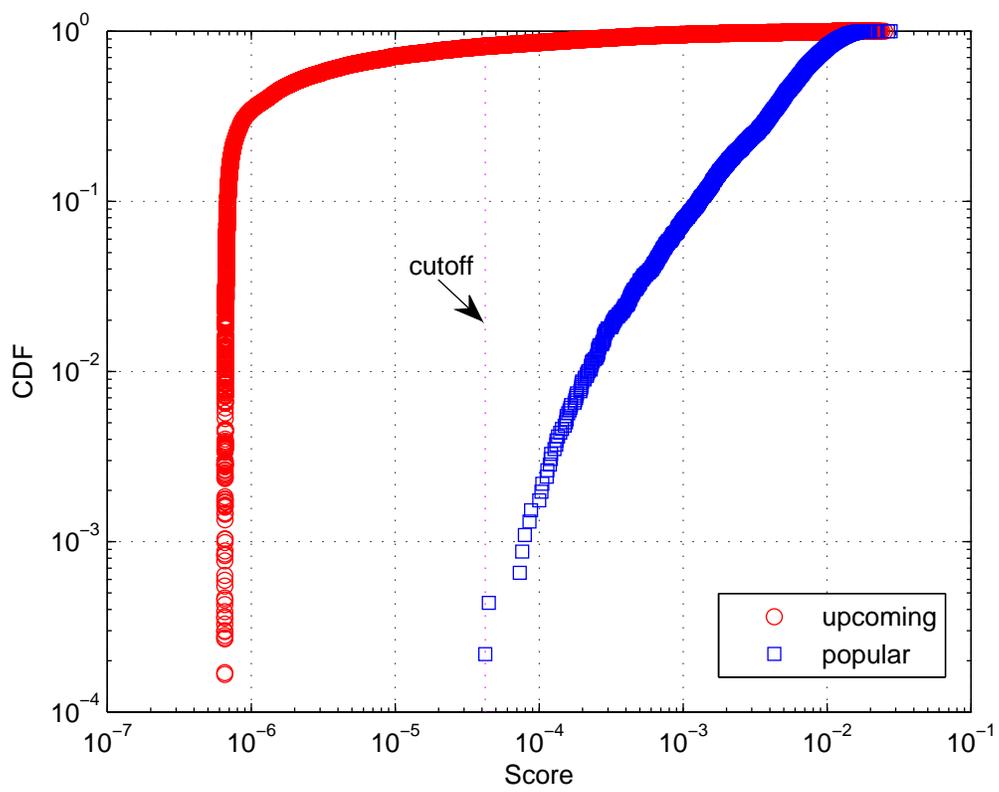